\documentstyle[prl,aps,epsfig,psfig,multicol]{revtex}

\def\eps{\varepsilon}

\begin{document}
\draft

\title{Scaling behavior in a stochastic self-gravitating system}

\bigskip

\author{N.V. Antonov$^{1,2}$}

\bigskip

\address{$^{1}$Department of Theoretical Physics, St.Petersburg University,
Uljanovskaja 1, St.Petersburg, Petrodvorez, 198504, Russia, \\
E-mail: Nikolai.Antonov@pobox.spbu.ru \\
$^{2}$Intelligent Systems Laboratory, Swedish Institute
of Computer Science, Box 1263, SE-164\,29 Kista, Sweden}
\bigskip

\maketitle

\bigskip

\centerline{Abstract}

\bigskip

\begin{abstract}
A system of stochastic differential equations for the velocity and density
of a classical self-gravitating matter is investigated by means of the field
theoretic renormalization group. The existence of two types of large-scale
scaling behavior, associated to physically admissible fixed points of the
renormalization-group equations, is established. Their regions of stability
are identified and the corresponding scaling dimensions are calculated in
the one-loop approximation (first order of the $\eps$ expansion).
The velocity and density fields have independent scaling dimensions.
Our analysis supports the importance of the rotational (nonpotential)
components of the velocity field in the formation of those scaling laws.

\end{abstract}

\bigskip

\begin{center}

SICS Technical Report T2003:09 \\
\bigskip
ISSN 1100-3154 \\
\bigskip
ISRN: SICS-T-2003/09-SE \\
\bigskip
Date: {\it 11 August 2003} \\
\end{center}

\begin{multicols}{2}

The present Universe at short and moderate distances is inhomogeneous,
being filled by numerous structures of many scales, from galaxies to galaxy
clusters and superclusters. On the contrary, at larger scales or earlier
stages the Universe is generally taken nearly homogeneous and isotropic.
It is believed that gravity reinforces small asymmetries in the velocity
and density fields, and the structures observed today are due to
instabilities in an initially uniform self-gravitating medium.
The first structures to form are ``pancakes,'' thin in one dimension and
of large extent in the two others. Further evolution and interaction of the
pancakes develops patterns with complex (fractal or honeycomb) geometry
in the distribution of matter.
Excellent reviews of classical cosmology are given in
Refs.~\cite{Zel,Wein,Peebles}.

The full relativistic treatment of the large-scale structure formation,
especially of its nonlinear stage, is an extremely difficult task. Therefore,
the development of instabilities is usually studied within the framework of
simplified dynamical models of a classical self-gravitating fluid (or system
of particles): Vlasov--Poisson model, adhesion model and its modifications
with different types of nonlinearity, pressure and viscous terms and random
forces, Boltzmann equation (or $N$-body simulations of dark matter) and so
on; see e.g. Refs.~\cite{Peebles,Frisch} for reviews and discussion.

The link between the complex geometrical structures and the coarse-grained
(hydrodynamic) description is provided by nontrivial scaling behavior
exhibited by correlation functions of the density or velocity fields, such as
the galaxy--galaxy correlation function $\xi(r)$; see e.g.~\cite{Japp}.
At large scales it reveals a power-law behavior
$\xi(r)\propto r^{-\gamma}$, where $\gamma$ is determined from catalogs
to be between 1.3 and 2.1 for $r$ of order of the Megaparsec;
see e.g. \cite{Pons,Norberg,MS,Zehavi,Zan} and references therein.
The most recent studies give the values between 1.6 and 1.9,
in particular, $\gamma=1.75\pm0.03$ according to \cite{Zehavi}.

The scope of theory is to derive such behavior on the basis of an
appropriate dynamical model, to investigate the universality of the exponent
$\gamma$, that is, its (in)dependence on the model parameters, and to
calculate $\gamma$ within a consistent approximation or systematic
perturbation scheme.

Scaling laws are typical of equilibrium phase transitions, and the most
adequate tool to study them is the renormalization group (RG). It is also
applicable to nonequilibrium dynamical phenomena as disparate as surface
growth, random walks, nonlinear diffusion and turbulence; see e.g.
\cite{Zinn} for a review. For a given model, the RG allows
one to prove the existence of
scaling regime(s), to determine the range of its stability in the space of
model parameters, and to calculate the scaling dimensions
in the form of regular perturbation series ($\varepsilon$ expansions).

The RG approach to the problem of self-gravitating medium was pioneered in
\cite{5,6,7}. In these studies, the full set of equations (hydrodynamic
equation for the Newtonian fluid, continuity equation and the Poisson
equation for the gravitation force, the system known as the Vlasov--Poisson
equations) was reduced to a single
equation for a purely potential velocity field.
The resulting equation (similar to the well-known stochastic Burgers equation
but with a time-dependent ``mass'' term) was augmented by a Gaussian random
force (noise) that represents the influence of fluctuations and dissipative
processes on the evolution of fluid, arising from viscosity, turbulence,
explosions, gravitational waves and so on. The dynamical RG approach of
\cite{FNS,KPZ,Medina} was then adopted to derive the scaling regimes and
exponents. With an appropriate choice of the parameters, the model reveals
scaling behavior with a nonuniversal exponent  $\gamma$; its value depends
on the characteristics of the forcing and can be adjusted to the value
$\gamma\approx 1.7$ \cite{7} in agreement with the observations.

In spite of this obvious success, the analysis of Refs. \cite{6,7} raises
serious questions about its internal consistency and interpretation
of the results. It is well known that the stochastic Burgers
(or Kardar--Parisi--Zhang) equation has no infrared-attractive fixed point
in the physical range of parameters within the $\varepsilon$ expansion.
This fact immediately follows from the first-order expressions of Refs.
\cite{FNS,KPZ,Medina}. It was confirmed by the two-loop calculation
\cite{10} and then proved to all orders of the perturbation theory \cite{11}.
The existence of a strong-coupling fixed point, although supported by
numerical simulations, remains an unproved hypothesis.

The authors of \cite{7} studied an extended (``massive'') version of
the KPZ model, and the only attractive fixed point revealed in their
analysis corresponds to nonzero value of the ``mass.'' In fact, its value
is comparable with the largest, ultraviolet, momentum scale of the problem,
so that the mass term at the fixed point is by no means small.
Therefore, in the spirit of the Landau theory, the viscous and
nonlinear terms in the equation should be discarded as infrared-irrelevant.
This leads to a simple Gaussian model whose critical exponents are easily
found exactly. They agree with the answers obtained in \cite{7} for the
full (interacting) model, but the situation on the whole is not quite
satisfactory. It looks rather strange that the nonlinearity does not play
any role in shaping the large-scale asymptotic form of the correlation
functions, and that the latter are so directly determined by the choice
of the random forcing. The absence of the appropriate fixed point for the
massless (critical) model suggests that such results can be very sensitive
to the approximations made in the derivation of the model. The most important
of them is the assumption of parallelism, which reduces the system of
equations for the density and velocity fields to an equation for a single,
purely potential, velocity field. More extensive analysis of the full problem
would therefore be desirable.

The original (deterministic) set of Vlasov--Poisson equations for
the velocity $u_{i}(t,{\bf x})$, density $\rho(t,{\bf x})$ and the
gravitational potential $\psi(t,{\bf x})$ in the comoving frame of
reference reads (see e.g. \cite{Peebles,Frisch}):
\begin{eqnarray}
a\partial_t u_{i} + \dot a u_{i} + (u_j\partial_{j})u_{i} -
\partial_{i} \psi=0,
\nonumber \\
a\partial_t \rho+ \partial_{i} (\rho u_{i} )=0,
\quad
\partial^{2} \psi = 4\pi Ga^{2} (\rho-\rho_{0}),
\label{1}
\end{eqnarray}
where $G$ is the gravitational constant, the cosmic scale factor $a(t)$ is
a prescribed function of the proper time, $\rho_{0}$ is the mean density and
$\partial^{2}\equiv \partial_{i}\partial_{i}$ is the Laplace operator.

We change to the new variables $\phi_{i}\equiv u_{i}/a$,
$\theta\equiv c^2 (\rho-\rho_{0})/ \rho_{0}$ and
$c^{2} \equiv 4\pi G a^{2}\rho_{0}$ and rewrite the system (\ref{1})
in the form:
 \end{multicols}
\begin{equation}
\left(\partial_{t} +  \phi_{j}\partial_{j} \right) \phi_{i} = -
H \phi_{i} + \nu_0 \left(\delta_{ij} \partial^{2}
-\partial_{i}\partial_{j} \right) \phi_{j} + u_{0}\nu_0
\partial_{i}\partial_{j} \phi_{j} - c\partial_{i}\partial^{-2} \theta
 +f_{i}, \
\partial_{t} \theta =  v_{0}\nu_0 \partial^{2} \theta
- \partial_{i} (\phi_{i}\theta)
-c (\partial_{i}\phi_{i}).
\label{2}
\end{equation}
 \begin{multicols}{2}
\noindent Here, $H\equiv \dot a/a$ is the Hubble function and $\partial^{-2}$
is the Green function of the Laplace operator. We have eliminated the
potential $\psi$ using the last equation in (\ref{1}), assumed that
$\dot a/a \ll \partial_{t} u/u$, and added viscous terms and the random force
$f_{i}(t,{\bf x})$. We stress that the velocity field $\phi_{i}\propto u_{i}$
is not purely potential, so that two independent viscosity coefficients
$\nu_0$ and $u_{0}\nu_0$ have been introduced in the equation for $\phi_{i}$.
The viscous term $v_{0}\nu_0 \partial^{2}\theta$ is usually not included in
the continuity equation, but it is not forbidden by dimensional reasons or
symmetry and thus is needed to ensure the renormalizability of the model.
Then the RG equations should be solved with the physical initial condition
$v_{0}=0$, but if the IR attractive fixed point is unique, the large-scale
behavior will be the same as for nonzero $v_{0}$; cf. the discussion
in~\cite{TMF}. Dimensional analysis shows that the pressure term is
infrared-irrelevant (in the sense of Wilson) in comparison to the
gravitational force and thus it was dropped in~(\ref{2}).

The hydrodynamic description of the properly smoothed (coarse-grained)
fields and, in particular, inclusion of the viscous terms and random
forcing, can be justified by various arguments \cite{Scope,Gidr,Quin}.
For simplicity, we shall neglect the time dependence of the viscosity
coefficients, suggested by those studies, treating it as a kind of
second-order
effect: the viscosity coefficients are ``small'' and their time dependence
is ``slow.'' In contrast to equilibrium systems, there is no universal
relation between the viscosity coefficients and the correlation functions
of the random force. We shall take it Gaussian, white in time (this is
necessary to ensure the Galilean symmetry of the stochastic problem
(\ref{2})), with zero mean and a given correlator
\begin{equation}
\big\langle f_{i}(t,{\bf k}) f_{j}(t',-{\bf k}) \big\rangle =
\delta(t-t') D(k) \left\{
P_{ij} + \alpha Q_{ij} \right\} .
\label{3}
\end{equation}
Here $P_{ij}=\delta_{ij}- k_i k_j/k^2$ and $Q_{ij}=k_i k_j/k^2$ are the
transverse and the longitudinal projectors, $\alpha>0$ is an arbitrary
parameter and $D(k)$ is a function of the modulus of the wave vector
$k\equiv|{\bf k}|$. The simplest possible choice is $D(k)=D_{0}$=const
(spatially decorrelated forcing). Another possibility, widely used in models
of nonequilibrium critical phenomena, is a power-law correlation function:
$D(k)=D_{0}' k^{4-d-2\eta}$; see e.g.
\cite{FNS,Medina,Dominicis,Frisch2,turbo,Two}. Here $d$ is the (arbitrary)
dimensionality of space and $\eta$ an arbitrary exponent; the notation is
explained by convenience reasons. In what follows, we shall refer to these
two cases as the local and nonlocal ones. They are closely related for
$d\le4$ (see below), but it is instructive to discuss them separately
in the beginning.

The RG analysis of a stochastic problem like (\ref{2}), (\ref{3}) includes
four important steps: field theoretic formulation; analysis of its
renormalizability; derivation of the corresponding RG equations; analysis
of the fixed points of these equations. This analysis for our problem is
technically involved (already in the simplest one-loop approximation) and
will be presented elsewhere, along with the details of the practical
calculation. In many respects, it is close to the field theoretic RG analysis
of the stochastic Navier-Stokes equation \cite{Dominicis,Frisch2,turbo,Two}
and especially to the case of a strongly compressible fluid studied in
\cite{TMF}. Below we only give the main points and conclusions.

According to a general theorem (see e.g. \cite{Zinn,turbo}), the stochastic
problem (\ref{2}) is equivalent to the field theoretic model of a doubled
set of fields $\Phi=\{\phi',\theta',\phi,\theta\}$ with action functional:
 \end{multicols}
\begin{eqnarray}
S(\Phi) = (1/2) \phi'D_{f}\phi' &+& \phi'_{i} \left\{- \left(\partial_{t}
+ \phi_{j}\partial_{j}+ H \right) \phi_{i}
+ \nu_0 \left(\delta_{ij} \partial^{2}
-\partial_{i}\partial_{j} \right) \phi_{j} + u_{0}\nu_0
\partial_{i}\partial_{j} \phi_{j} - c \partial_{i}\partial^{-2}
\theta \right\}
\nonumber
\\
&+& \theta' \left\{ -\partial_{t} \theta
+ v_{0}\nu_0 \partial^{2} \theta - \partial_{i} (\phi_{i}\theta)
-c (\partial_{i}\phi_{i}) \right\} ,
\label{action}
\end{eqnarray}
 \begin{multicols}{2}
\noindent where $D_{f}$ is the correlator (\ref{3}) and the needed
integrations over $t,{\bf x}$ and summations over the vector indices
are implied.

The field theoretic formulation means that the correlation functions of the
stochastic problem (\ref{2}), (\ref{3}) can be represented as functional
averages with the weight $\exp S(\Phi)$ with action (\ref{action}). This
allows one to use a well-developed formalism (power counting plus symmetries
of the model) to analyze the relation between the IR and UV problems and the
UV renormalizability of the model. For the
local case, it shows that the upper critical dimension for the model is
$d=4$: the nonlinearity in (\ref{2}) is IR irrelevant for $d>4$ (perturbation
theory is applicable, no scaling and universality are expected). For $d\le4$,
the terms of the ordinary perturbation theory suffer from IR singularities
and cannot be used to describe the large-scale behavior of the problem.
For small $\varepsilon\equiv d-4$, the problem of the IR singularities
is closely related to that of the UV divergences (poles in $\varepsilon$).
The latter is solved by the standard UV renormalization procedure: it shows
that the model (\ref{action}) is multiplicatively renormalizable, that is,
all the poles in $\varepsilon$ in its correlation functions are removed by
the rescaling of the fields $\Phi$ and the parameters
$D_{0},\nu_0,u_{0},v_{0},\alpha$ (the proof of this statement is the most
nontrivial stage of the analysis). The arbitrariness in the renormalization
procedure leads to the RG equations: first-order differential equations for
the correlation functions with coefficients calculated within the ordinary
perturbation theory. In order to draw any definite conclusions from the RG
equations, one has to calculate their coefficients at least in the simplest
(one-loop) approximation. We performed the calculation and found out that,
in contrast to the Burgers or KPZ models, the RG equations of the extended
model (\ref{action}) have the only IR attractive fixed point in the
physical range of parameters (the ratios of the viscosity coefficients
and the amplitude factors in pair correlation functions are positive).

This means, in particular, that in the IR range (the scales large in
comparison to the typical UV length scale, built of $D_0$ and $\nu_0$,
and times large in comparison to the corresponding time scale), the
correlation functions of the velocity $\phi$ and the density (more
precisely, of the field $\theta\equiv c^{2} (\rho-\rho_{0})/ \rho_{0}$
have a scaling (self-similar) form:
\begin{eqnarray}
\big\langle \phi(t,{\bf x}) \phi(t+\tau,{\bf x}+{\bf r})\big\rangle
\simeq r^{-2\Delta_{\phi}}\, F_{\phi}(\dots)\, ,
\nonumber \\
\big\langle \theta(t,{\bf x}) \theta(t+\tau,{\bf x}+{\bf r})\big\rangle
\simeq r^{-2\Delta_{\theta}}\, F_{\theta}(\dots)\, ,
\label{CF}
\end{eqnarray}
where $r\equiv |{\bf r}|$ and the scaling functions $F_{v,\theta}$ depend on
(critically) dimensionless variables $\tau\cdot r^{\Delta_{\tau}}$,
$H\cdot r^{\Delta_{H}}$, $c\cdot r^{\Delta_c}$
(for the equal-time correlation functions, the first variable is absent).
The dimensions $\Delta$ are universal in the sense that
they are independent of the values of the parameters $u_{0},v_{0},\alpha$
and can be calculated as series in $\varepsilon$. The first-order (one-loop)
calculation gives:
\begin{eqnarray}
\Delta_{\phi}&=&1-\eps/2, \quad  \Delta_{\theta}=\Delta_{c}=2-\eps/2,
\nonumber \\
\Delta_{\tau}&=&-2+\eps/2, \quad \Delta_{H}= 2+ \eps/2,
\label{Deltas}
\end{eqnarray}
with corrections of order $\eps^{2}$ and higher.
From representations (\ref{CF}) it follows that under the rescaling
\begin{eqnarray}
r\to r/\Lambda, \quad
\tau \to \tau \Lambda^{\Delta_{\tau}}, \quad
H \to H \Lambda^{\Delta_{H}}, \quad
c \to c \Lambda^{\Delta_{c}}
\label{Lamb}
\end{eqnarray}
with arbitrary $\Lambda>0$, the correlation functions behave as
\begin{eqnarray}
\langle v v\rangle \to  \Lambda ^{2\Delta_{v}} \langle v v\rangle,
\qquad  \langle \theta \theta \rangle  \to  \Lambda
^{2\Delta_{\theta}} \langle \theta \theta \rangle .
\label{CFF}
\end{eqnarray}
The formulation (\ref{Lamb}), (\ref{CFF}) is in fact more general because
it remains true if the parameters $H,c$ depend on $t,{\bf x}$ (in the
original problem they indeed depend on $t$), while the more explicit formulae
(\ref{CF}) imply that they are treated as constants.

The RG representations (\ref{CF}) are the result of certain infinite
resummation of the primitive perturbation theory, that is, of the
expansion in the nonlinearity in Eqs. (\ref{2}) around the zero-order
(Gaussian) approximation. In our case, however, the latter is unstable with
respect to any small perturbation, as is easily seen from the fact that,
for $c^{2}>0$, the retarded zero-order response function grows in time and,
as a result, perturbative diagrams contain infrared divergences. However,
it can be argued that this instability does not hinder the use of the
RG in studying the self-similar behavior. The parameters
$H$ and $c$ in (\ref{action}) have integer positive dimensions and,
in this respect, they are analogous to masses (in the language of the
quantum field theory) or to the deviation of the temperature of its
critical value, $\Delta T\equiv T-T_{c}$ (in models of critical behavior).
From the general theory of UV renormalization, it is well known that the
UV divergent parts of the diagrams are polynomials in such ``IR relevant
parameters.'' Therefore, they can
be calculated for $\Delta T \ge 0$ (or $c^{2}\le0$ in our case), where
the terms of the perturbation theory are finite, and then extrapolated to
the region $\Delta T<0$ (or $c^{2}>0$). There, the ordinary perturbation
expansion ceases to make sense due to IR divergences and one should either
change to an improved perturbation theory (for critical behavior) or to
consider a nonstationary problem (for a self-gravitating system). It is
important here that this rearrangement does not affect the UV divergent
parts of the correlation functions (and hence the counterterms).
These arguments show that the critical exponents (calculated from the
UV counterterms) are the same below and above $T_{c}$, and, in our case,
they support the scaling relations (\ref{CF}) and (\ref{Lamb}), (\ref{CFF})
with the dimensions (\ref{Deltas}) for the ``unstable'' case $c^{2}>0$.

For the nonlocal case, that is, $D(k)\propto k^{4-d-2\eta}$ in (\ref{3}),
analysis shows that the model (\ref{action}) appears multiplicatively
renormalizable if $d>4$, and the corresponding RG equations also have an IR
attractive fixed point in the physical range of the parameters. This
establishes the scaling relations (\ref{CF}), (\ref{Lamb}), (\ref{CFF})
with the new set of dimensions:
\begin{eqnarray}
\Delta_{\phi}&=& 1-2\eta/3\ ({\rm exact}), \quad
\Delta_{\theta}=\Delta_{c}=2-2\eta/3, \nonumber \\
\Delta_{\tau}&=&-2+2\eta/3, \quad \Delta_{H}= 2+ 2\eta/3 \ ({\rm exact}).
\label{DeltaZ}
\end{eqnarray}
The dimensions $\Delta_{\phi,H}$ are found exactly (there are no corrections
of order $\eta^{2}$ and higher) due to the Galilean invariance of the problem.
In principle, the other dimensions are less universal than their analogues
for the local case: besides the exponent $\eta$, they can depend on
$d$ and $\alpha$ from (\ref{3}). Our calculation has shown, however,
that this dependence can occur only in the order $O(\eta^{2})$.

For $d\le4$, the model (\ref{action}) with the nonlocal noise correlator
ceases to be renormalizable: a new counterterm $(\phi')^2$ is generated.
A similar problem is well known in the RG approach to the stochastic
Navier--Stokes equation for purely incompressible fluid, where it occurs
at $d=2$; see e.g. Sec.~3.10 of Ref. \cite{turbo}.
In order to apply the RG to this case, one has to extend the model by adding
such term to the action from the very beginning, that is, one has to study
the model with the mixed correlator
\begin{eqnarray}
D(k) = D_{0} + D_{0}'\,k^{4-d-2\eta}
\label{Mix}
\end{eqnarray}
(similar to that discussed in \cite{6,7,Medina}). The extended model appears
renormalizable, and its fixed points can be studied within the double
expansion in two parameters, $\eta$ and $\varepsilon=4-d$.

The calculation in the first-order of such expansion shows that the extended
model has two nontrivial fixed points. The first of them is IR attractive for
$\varepsilon>0$, $\eta<3\varepsilon/4$ and corresponds to the ``local''
regime with the dimensions (\ref{Deltas}), while the second is IR attractive
for $\eta>0$, $\eta>3\varepsilon/4$ and corresponds to the dimensions
(\ref{DeltaZ}). For $\eta<0$, $\varepsilon<0$ the only IR attractive point
is trivial; it corresponds to a free (non-interacting) field theory.
The regions of stability of the fixed points of the extended model in
the $\varepsilon$--$\eta$ plane are shown in Fig.~1.

The main conclusions of our analysis are as follows. We have investigated a
system of stochastic differential equations for the velocity and density of
a self-gravitating matter, established two types of large-scale scaling
behavior (local and nonlocal ones), identified their regions of stability
and calculated the scaling dimensions in the one-loop approximation (i.e.,
to first order of the corresponding $\eps$ expansions).

From the qualitative point of view, our analysis shows that nonequilibrium
stochastic systems of the type (\ref{2}) can have IR attractive fixed points
in the physical range of parameters, and the corresponding scaling regimes
can be treated systematically, within appropriate $\eps$ expansions.
What is more, such models can have several fixed points with different sets
of dimensions, and the system undergoes the crossover in its large-scale
behavior when its parameters (exponents in the forcing) change.

It is worth noting that in model (\ref{2}), the density and velocity fields
have independent scaling dimensions, a feature which is lost if the full set
of equations is reduced to a single equation for only one independent field.
Our results also suggest that rotational (non-potential) components of the
velocity field do not decouple in those regimes and should be taken into
account in the analysis of the large-scale behavior.

Admittedly, the one-loop answers for the exponents are markedly larger than
the latest experimental estimates for the exponent $\gamma$ (identified
with $2\Delta_{\theta}$). In particular, for the local regime and $d=3$ one
obtains $\gamma\approx3$. For the nonlocal regime and
arbitrary spatial dimension, $\gamma$ varies from 4 to 2 when the exponent
$\eta$ varies within its natural range $0<\eta<3/2$ (for $\eta>3/2$, the
dimensions $\Delta_{\phi,\tau}$ become negative).
This can be a hint that the simplified model (\ref{2}) does not include all
physical interactions relevant for the large-scale behavior (it is worth
mentioning here that the model (\ref{1}) concerns directly dark matter,
while the galaxy-galaxy correlation functions concern visible matter).

One can also expect that the simplest one-loop
approximations (\ref{Deltas}), (\ref{DeltaZ}) overestimate the value of
the scaling dimensions, and the second-order and higher corrections will
improve the agreement, as indeed happens in the RG theory of fully
developed turbulence; see \cite{Two}. Finally, it is possible that the
scaling functions $F$ in representations (\ref{CF}) are very singular in
their arguments, which can lead to imaginary shift of the genuine
exponent or to deviation from a plain power-law behavior, in agreement
with some recent data \cite{Devi}.

In order to investigate these issues, one should go beyond the simplest
one-loop approximations and augment the plain RG equations by more advanced
tools (renormalization of composite operators, operator-product expansion
and so on), in analogy with the RG theory of fully developed turbulence;
see e.g. \cite{turbo}. This work is left for the future.

I thank Erik Aurell and Paolo Muratore Ginanneschi for numerous discussions
and valuable suggestions.
This work was performed within the framework of the Visby Program offered by
the Swedish Institute. It was also supported in part by the Nordic Grant for
Network Cooperation with the Baltic Countries and Northwest Russia
Nos.~FIN-6/2002 and~FIN-20/2003, the Academy of Finland (Grant No.~203122)
and the program ``Universities of Russia.'' I thank the Swedish Institute
of Computer Science in Kista for their kind hospitality.

\end{multicols}

\newpage

\begin{figure}
\centerline{
\epsfig{file=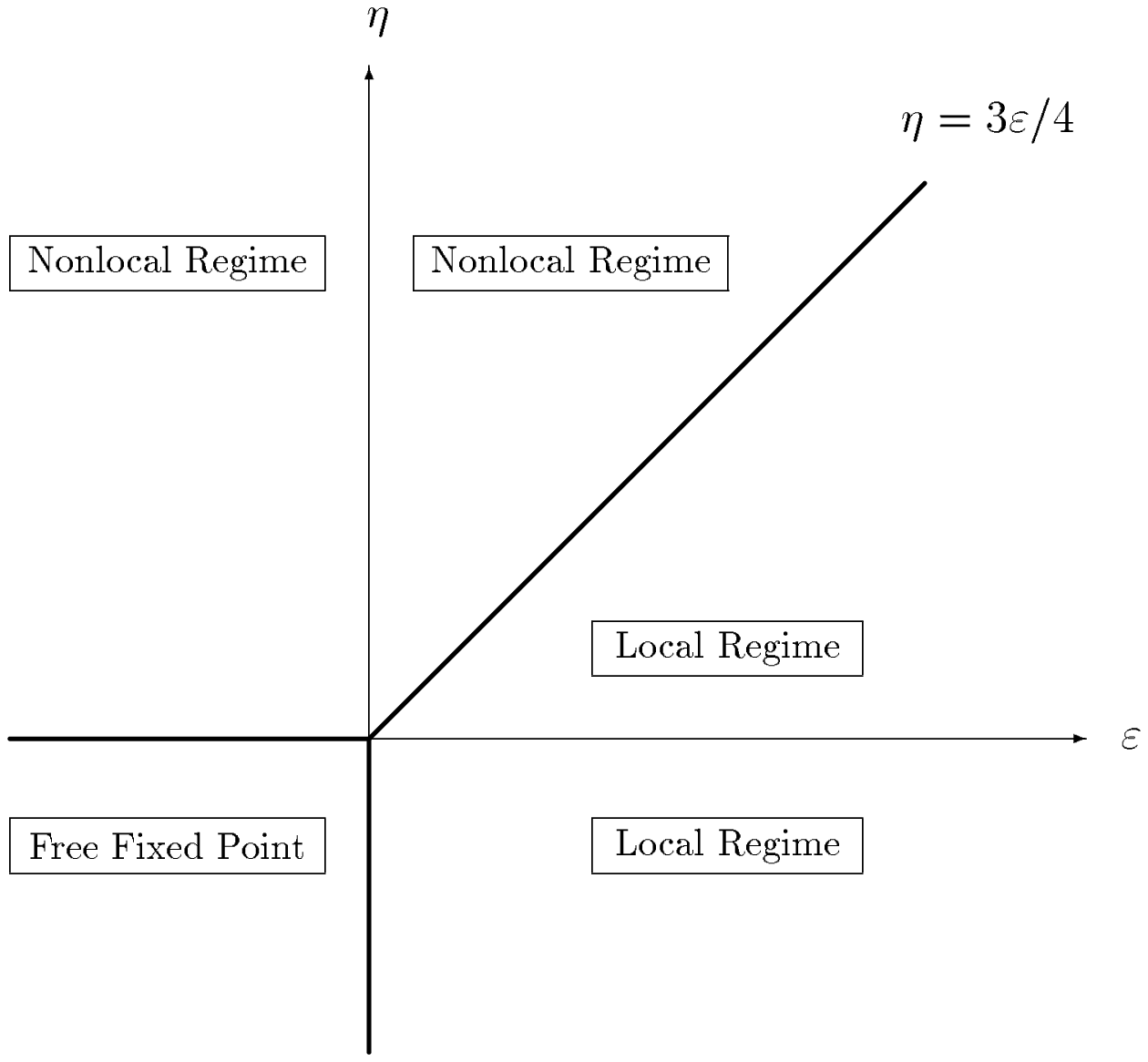,width=15cm}}
\end{figure}

\vspace{-7cm}
{\small FIG.~1. Phase diagram of the model (\protect\ref{action}),
(\protect\ref{Mix}),
in the $\varepsilon$--$\eta$ plane: the local regime is realized for
$\varepsilon>0$, $\eta<3\varepsilon/4$, the nonlocal one for $\eta>0$,
$\eta>3\varepsilon/4$ and the trivial one for $\eta$, $\varepsilon<0$.}

\end{document}